# Mutual Coupling Reduction in Two-Dimensional Array of Microstrip Antennas Using Concave Rectangular Patches

Shahram Mohanna, Ali Farahbakhsh, and Saeed Tavakoli

**Abstract**— Using concave rectangular patches, a new solution to reduce mutual coupling and return loss in two-dimensional array of microstrip antennas is proposed. The effect of width and length concavity on mutual coupling and return loss is studied. Also, the patch parameters as well as the amounts of width and length concavity are optimized using an enhanced genetic algorithm. Simulation results show that the resulting array antenna has low amounts of mutual coupling and return loss.

———————————— ◆ ————————————

## 1 INTRODUCTION

MICROSTRIP array antennas are used widely because of their simple manufacturing, small size, light weight and low cost [1-6]. They are used in phased array antennas applications such as pattern beam forming, smart antennas, and electronic scanning radars [4, 5]. To calculate the radiation pattern of an array the mutual coupling effect must be considered. Without considering this effect, large errors in beam forming and null string are resulted [7, 8]. To decrease the mutual coupling effect, several methods such as changing feed position and feed structure and replacing ordinary patches with new types, such as fractal patches, have been reported [9-14].

To reduce mutual coupling, the use of concave rectangular patches is proposed. In [9], this method was applied to a linear array. In this paper, the patch concavity effect on a two-dimensional array, including four elements, is studied. In addition, the effect of patch concavity on return loss is investigated. Furthermore, the patch length and width as well as the amount of width and length concavity are optimized employing an enhanced genetic algorithm.

## 2 STRUCTURE OF THE ARRAY ANTENNA

The array antenna includes four elements, which all of them are implemented on one antenna layer structure. The antenna layer structure includes three conductive

————————————————

• S. Mohanna is with the Faculty of Electrical and Computer Engineering, The University of Sistan and Baluchestan, Zahedan, Iran.
• A. Farahbakhsh is with the Faculty of Electrical and Computer Engineering, The University of Sistan and Baluchestan, Zahedan, Iran.
• S. Tavakoli is with the Faculty of Electrical and Computer Engineering, The University of Sistan and Baluchestan, Zahedan, Iran.

and two dielectric layers. The lower conductive layer is assumed to be an infinite ground plane, while the second one is set to be a feed layer and the patches are placed in the top layer. In simulation, the thickness of dielectric layers, $D_1$ and $D_2$, are set to 0.5mm. The dielectric relative permittivity, $\varepsilon_r$, and the dielectric relative permeability, $\mu_r$, are equal to 3 and 1, respectively. Figure 1 shows the side view of the antenna.

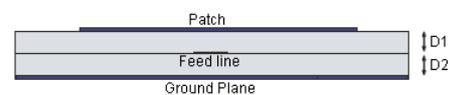

Fig. 1. Layer structure of the antenna

Figure 2 illustrates the array structure and its dimensions. The patch dimensions can be specified by the frequency and the bandwidth of the antenna [6]. In this study, the antenna is designed for X band applications. The patch width and length is set to be $W = 11.55mm$ and $L = 9.45mm$, results the resonant frequency to be 8.55GHz. The patches are fed by proximity coupled mechanism with L shape feed lines, having a line impedance of 50 Ohm. As shown in Figure 2, the distances between center of elements in X and Y directions are 14.08mm and 20.55mm, respectively.

## 3 PATCH CONCAVITY EFFECT

Figure 3 illustrates the concavity on the above-mentioned array antenna, where $h_1$ and $h_2$ refer to the depths of the width concavity and length concavity, respectively. In this section, the effect of concavity in width, length, and both sides on the mutual coupling and reflection coefficient is studied. Employing FEKO software [15], antenna parameters are obtained based on the method of moments [16].

## 3.1 Effect of Width Concavity

The effect of the width concavity of the patches ($h_1$) on the mutual coupling and reflection coefficient is investigated. The mutual coupling and reflection coefficient are computed for $h_1 = 0, 0.25, 0.5, 0.75, 1, 1.25 mm$.

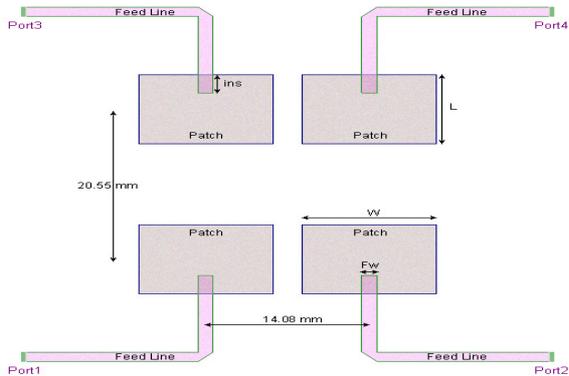

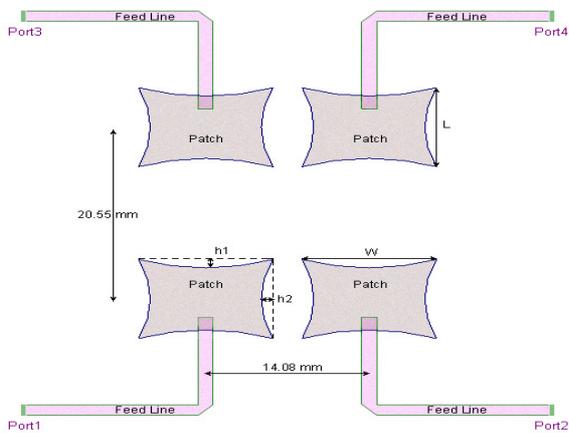

Fig. 2. Structure of array antenna used in the simulation

Fig. 3. Structure of concave array antenna

The patch with no concavity is shown by $h_1 = 0mm$. The scattering parameters, which are related to either the reflection coefficient or mutual coupling, are depicted in Figures 4 and 5. It can be seen that the reflection coefficient is increased and the mutual coupling is decreased by increasing the amount of concavity. Although the width concavity can decrease the effect of mutual coupling, the resonant frequency is shifted up. To compensate this effect and move the resonant frequency back to its initial value, patch dimensions should be increased.

## 3.2 Effect of Length Concavity

The effect of the length concavity of the patches ($h_2$) on mutual coupling and reflection coefficient is studied. Figures 6 and Figure 7 show the scattering parameters. It is seen that the reflection coefficient is minimized in $h_2 = 1mm$. By increasing the amount of the length concavity, parameters $S_{12}$ and $S_{13}$ are decreased, whereas $S_{14}$ is increased. Although the width concavity can decrease the effect of mutual coupling, the resonant frequency is shifted down. To move the resonant frequency back to its initial value, patch dimensions should be decreased. This effect is useful in designing small-size antennas.

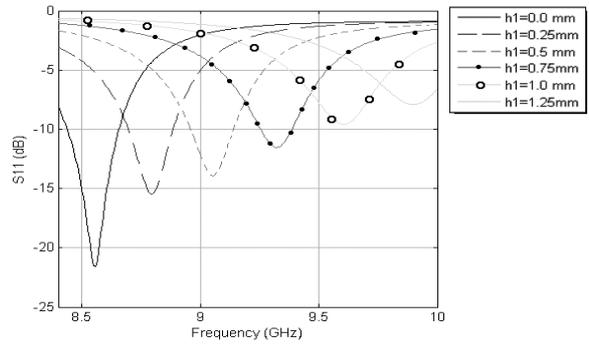

Fig. 4. Reflection coefficient for width concavity

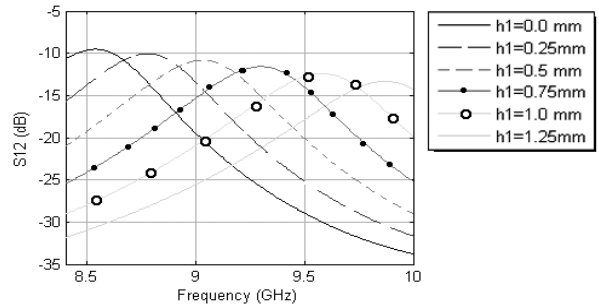

(a)

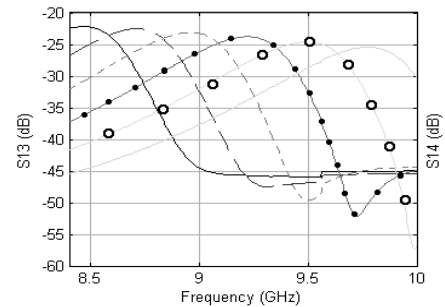

(b)

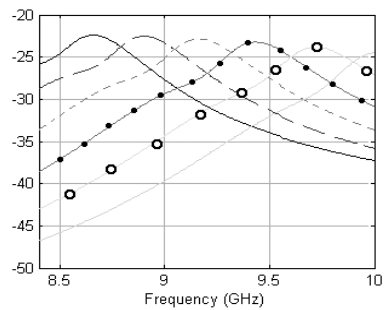

(c)

Fig. 5. Coefficients related to mutual coupling for width concavity

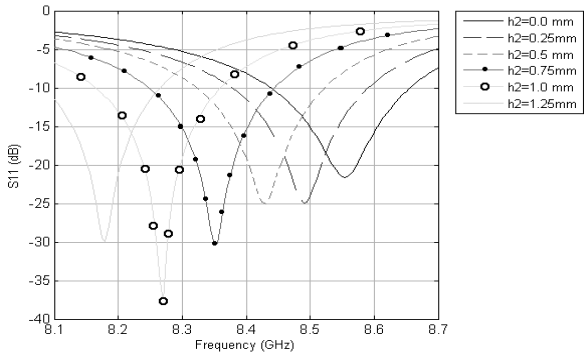

Fig. 6. Reflection coefficient for length concavity

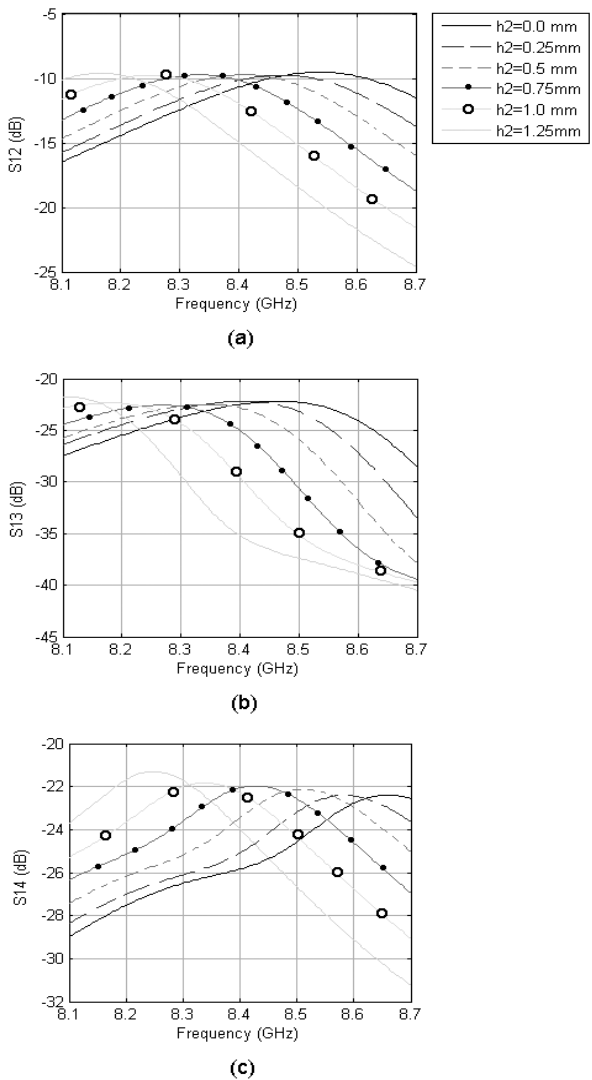

Fig. 7. Coefficients related to mutual coupling for length concavity

## 3.3 Effect of Two- Side Concavity

For simplicity the depths of concavity in both sides, $h_1$ and $h_2$, are considered to be the same. The effect of the two-side concavity on the reflection coefficient and mutual coupling are shown in Figures 8 and 9. By increasing the amount of concavity, it can be seen that the reflection coefficient is increased, whereas the mutual coupling is decreased. Like the previous cases, the resonant frequency is changed.

From all three cases considered in this section, it can be concluded that the concavity can influence the mutual coupling effect and reflection coefficient. Therefore, by changing the size of width and length as well as the amount of concavity, an array antenna with a desired resonant frequency and low amounts of mutual coupling and reflection coefficient can be designed. In addition, the parameter showing the amount of feed line and patch overlap affects antenna properties. Therefore, this parameter should also be considered in the design procedure.

As this procedure is based on trial and error, it is time-consuming. More importantly, if the concavity in each side is considered to be different, the trial and error procedure will be more complicated. Therefore, it is proposed to employ an optimization procedure to determine the optimal values of the patch length, the patch width, the amount of overlap, as well as the amounts of width and length concavity.

## 4 ENHANCED GENETIC ALGORITHM

Based on an analogy to the phenomenon of natural selection in biology, Holland [17] proposed the GA method. First, the optimization problem is given a chromosome structure. Next, an initial population is generated, randomly. Then, members of the population with higher fitness are selected. The fitness of members is calculated by an evaluation function. A member with a higher fitness has a more chance to be selected; therefore, weaker members are gradually replaced by stronger members. Selected members mate two by two randomly and the next population is generated. This procedure is repeated until the stop condition is reached. To prevent the algorithm to converge to local optimums, the mutation operator, which generates new chromosomes with different characteristics, is also applied. The mutation and crossover operators can be interpreted as negative and positive feedbacks, respectively.

The coefficient of crossover acts as a positive feedback. When crossover coefficient is increased, the algorithm is forced to converge. If it is increased too much, the algorithm may converge to a local optimum. The mutation coefficient plays the role of the negative feedback. Increasing mutation coefficient results in a deep search but at the cost of a low-speed convergence.

In contrast to conventional GA, in which coefficients of feedbacks are constant during running the optimization process, the method presented in [18] suggests changing these coefficients, using fuzzy systems. This leads to a good trade-off between the speed of algorithm and the accuracy of the solution.

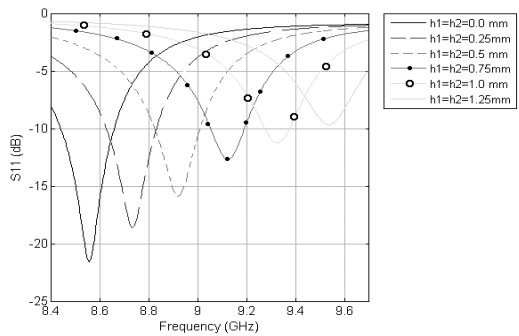

Fig. 8. Reflection coefficient for two-side concavity

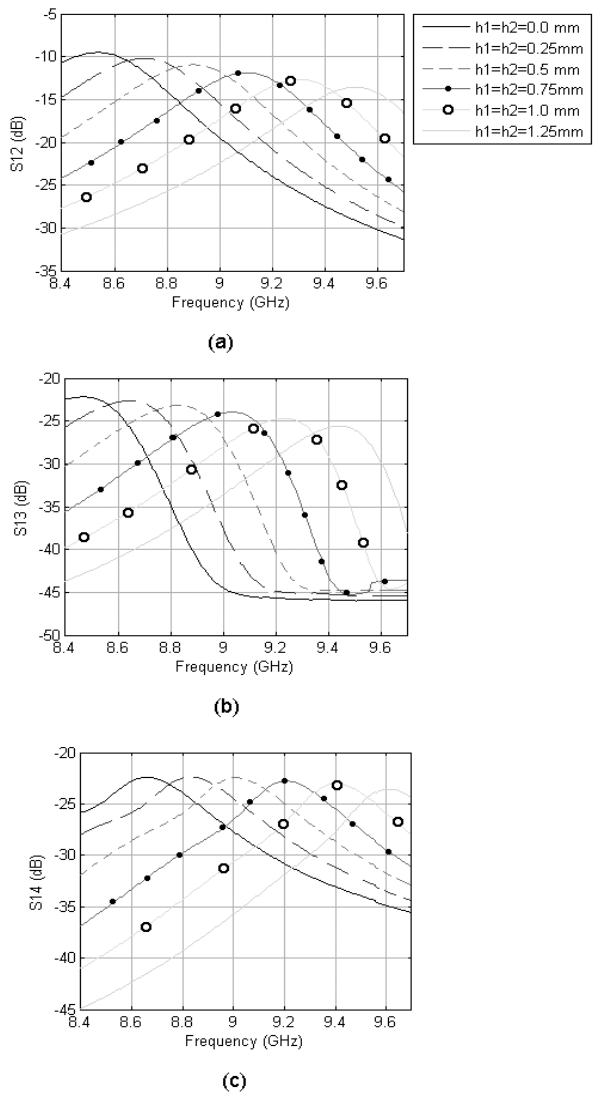

**(a)**

**(b)**

**(c)**

Fig. 9. Coefficients related to mutual coupling for two-side concavity

To increase the convergence of the algorithm, the positive feedback is increased at the start of the optimization process. Then, the negative feedback is increased to ensure the globality of the solution. Computing the convergence of the algorithm, coefficients of feedbacks are changed appropriately. When the algorithm is about to converge, therefore, the negative feedback should be increased. The positive feedback must be increased, when the algorithm fails to converge.

## 5 OPTIMIZATION OF ARRAY ANTENNA

To achieve an array antenna with the least amounts of mutual coupling and reflection coefficient, an enhanced genetic algorithm [18] is employed. The optimization procedure aims to minimize the scattering parameters at a resonant frequency of 8.55 GHz. Optimization variables are dimensions of the patches, $W, L$, the amount of feed line and patch overlap, $ins$, and the concavity parameters, $h_1, h_2$. Based on the above-mentioned optimization procedure, a MATLAB [19] code is developed, which is linked with FEKO. Running this program, the optimal parameters are given by $W = 9.213 mm$, $L = 10.103 mm$, $ins = 2.531 mm$, $h_1 = 1.527 mm$, and $h_2 = 1.601 mm$. The reflection coefficient and mutual coupling are shown in Figures 10 and 11.

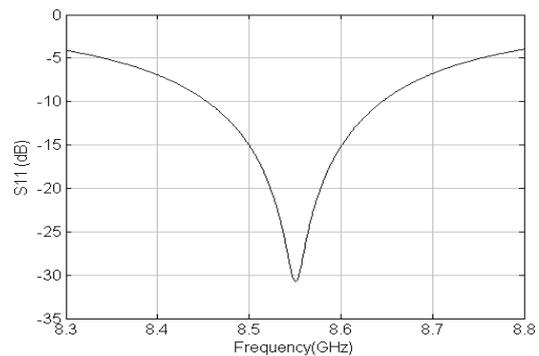

Fig. 10. Reflection coefficient for the optimal array antenna

## 6 CONCLUSIONS

This paper proposed a new solution to reduce mutual coupling in two-dimensional array of microstrip antennas by using concave patches instead of common rectangular patches. The effect of width and length concavity on the antenna mutual coupling and return loss was studied. Also, an optimal array antenna was designed by employing an enhanced genetic algorithm. Simulation results demonstrated that the resulting array antenna has low amounts of mutual coupling and return loss.

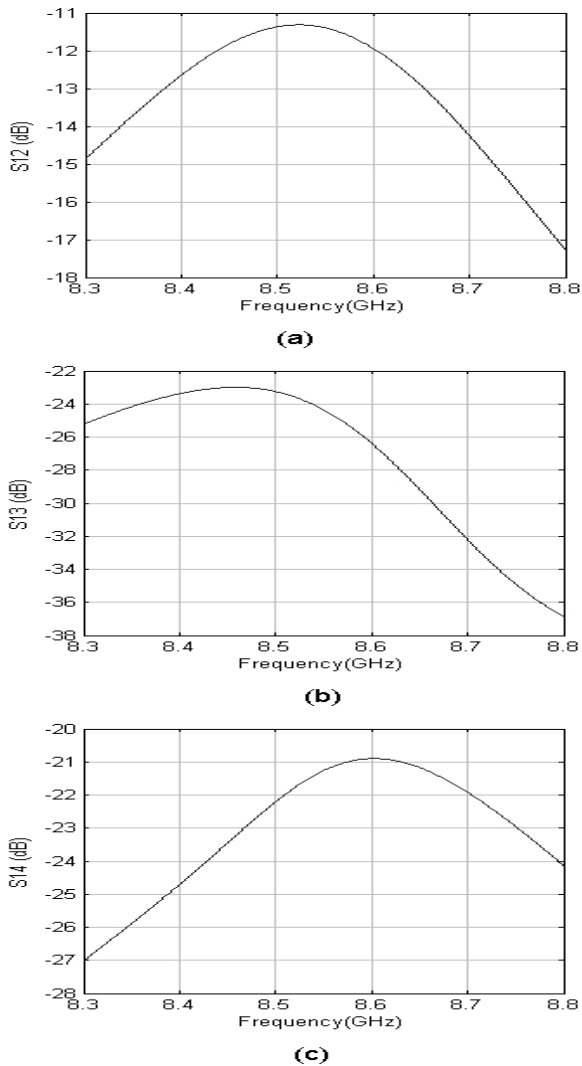

Fig. 11. Coefficients related to mutual coupling for the optimal array antenna

**Shahram Mohanna** received his BSc and MSc degrees in electrical engineering from the University of Sistan and Baluchestan, Iran and the University of Shiraz, Iran in 1990 and 1994, respectively. He then joined the University of Sistan and Baluchestan, Iran. In 2005, he obtained his PhD degree in electrical engineering from the University of Manchester, England. As an assistant professor at the University of Sistan and Baluchestan, his areas of research include design of microwave circuits, antenna design and applied electromagnetic. Dr. Mohanna has served as a reviewer for several journals and a number of conferences.

**Ali Farahbakhsh** obtained his BSc and MSc degrees in electrical engineering from Shahid Bahonar University of Kerman, Iran and the University of Sistan and Baluchestan, Iran in 2007 and 2010, respectively. His main area of research is design of microstrip array antennas.

**Saeed Tavakoli** received his BSc and MSc degrees in electrical engineering from Ferdowsi University of Mashhad, Iran in 1991 and 1995, respectively. In 1995, he joined the University of Sistan and Baluchestan, Iran. He earned his PhD degree in electrical engineering from the University of Sheffield, England in 2005. As an assistant professor at the University of Sistan and Baluchestan, his research interests are space mapping optimization, multi-objective optimization, control of time delay systems, PID control design, robust control, and jet engine control. Dr. Tavakoli has served as a reviewer for several journals including IEEE Transactions on Automatic Control, IEEE Transactions on Control Systems Technology, IET Control Theory & Applications, and a number of international conferences.